\documentstyle[aps,prc,psfig,floats,twocolumn]{revtex}

\tighten

\begin{document}

\preprint{}
\draft 

\wideabs{

\title{Gauge-invariant tree-level photoproduction amplitudes with form factors}

\author{H. Haberzettl,$^a$  C. Bennhold,$^a$ T. Mart,$^{a,b}$ and T. Feuster$^{c}$}

\address{$^a$Center for Nuclear Studies, Department of Physics, 
The George Washington University, Washington, D.C. 20052\\
$^b$Jurusan Fisika, FMIPA, Universitas Indonesia, Depok 16424, Indonesia\\
$^c$Institut f\"ur Theoretische Physik, Universit\"at Gie{\ss}en,
D-35392 Gie{\ss}en, Germany}

\date{\today}

\maketitle

\begin{abstract}  
We show how the gauge-invariance formulation given by Haberzettl is implemented in practice for 
photoproduction amplitudes at the tree level with form factors describing composite nucleons. 
We demonstrate that, in contrast to Ohta's gauge-invariance prescription, this formalism
allows electric current contributions to be multiplied by a form factor, i.e., it does not 
require that they be treated like bare currents. While different in detail, 
this nevertheless lends support to previous {\it ad hoc} approaches which multiply the Born 
amplitudes by an overall form factor. Numerical results for kaon photoproduction 
off the nucleon are given. They show that the gauge procedure by Haberzettl leads to 
much improved $\chi^2$ values as compared to Ohta's  prescription.

\end{abstract}

\pacs{PACS number(s): 25.20.Lj, 13.60.Le, 11.40.-q, 11.80.Cr} 
\vspace{-2mm}

}


The question of gauge invariance is one of the central issues
in dynamical descriptions of how photons interact with 
hadronic systems \cite{gaugeauthors,ohta89,WNP,hh97g}. While there is usually 
no problem to find definitive 
answers at the level of tree diagrams with bare, point-like particles, the 
problem becomes rapidly very complicated once one attempts to incorporate 
the electromagnetic interaction consistently within the
full complexity of a strongly-interacting hadronic system \cite{hh97g}. 
As a case in point, as is well known, even the tree-level amplitude
for pion photoproduction off the nucleon is not gauge-invariant if one 
employs hadronic $\pi NN$ form factors to account for the fact that 
nucleons are composite objects, and not point-like. 

In order to restore gauge invariance in these situations,
one needs to construct additional current contributions 
beyond the usual Feynman diagrams to cancel the gauge-violating terms.
One of the most widely used methods to this end is
due to Ohta \cite{ohta89}. For pion photoproduction off the nucleon at the 
level of the Born amplitude, Ohta's prescription amounts to dropping all strong-interaction 
form factors for all gauge-violating electric current contributions \cite{WNP}. 
In other words, gauge invariance is regained by treating the offending terms 
exactly as in the bare case, thus losing any effect due to the compositeness of the
nucleons. This  undesireable situation is sometimes remedied in an {\it ad hoc} 
fashion by multiplying the gauge-invariant bare amplitude by an overall form factor 
taken to simulate the average effect of the fact that nucleons are not point-like \cite{fmult}. 
Within Ohta's scheme, however, there is no foundation for such recipes \cite{WNP}.

Recently, Haberzettl \cite{hh97g} has put forward a comprehensive treatment of gauge 
invariance in meson photoproduction. This includes a prescription for restoring gauge 
invariance in situations when one cannot, for whatever reason, handle the full complexity of 
the problem and therefore must resort to some approximations. It is the purpose of the 
present paper to provide a detailed comparison of this approach with Ohta's. While the 
general {\it Ansatz} in Ref.\ \cite{hh97g} was quite different from Ohta's, we will show that 
both approaches can be understood as different ways of taking the limit of 
vanishing photon momentum. The way this limit is treated in Ref.\ \cite{hh97g}  will be seen to 
introduce more flexibility in how form factors can be retained for the terms where they 
are replaced by constants in Ohta's prescription. Although different in detail,
this finding actually lends support to approaches which multiply the Born amplitude by an 
overall form factor. 

We will use the reaction
$\gamma p \rightarrow n \pi^+$ with pseudoscalar coupling for the $\pi NN$ vertex as 
a simple example to elucidate the main features of the present investigation, similar 
to the discussion of Ohta's approach \cite{ohta89} in Ref.\ \cite{WNP}. Using different,
or more general, couplings for the vertex would not add anything essential to the following 
discussion; it would only complicate the presentation.

\begin{figure}[b!]%
\centerline{\psfig{file=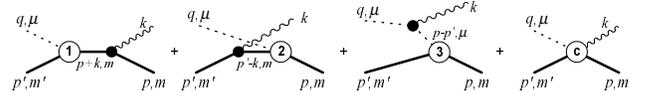,width=.95\columnwidth,clip=,silent=}} 
\vspace{2.5mm}
\caption{\label{fig1} Tree-level photoproduction diagrams. Time proceeds
from right to left. 
The form factors $F_1$, $F_2$, and $F_3$ in the text describe
the vertices labeled by 1, 2, and 3, respectively, with appropriate
momenta and masses shown for their legs. The right-most diagram
corresponds to the contact term $M_{\text{c}}^\mu$ required to restore gauge invariance
[Eq.\ (\ref{cc})]; it is
absent for pure pseudoscalar coupling with bare vertices.}%
\end{figure}

For bare nucleons, the tree-level amplitude (see Fig.\ \ref{fig1})%
%
%
%
for $\gamma  p \rightarrow n  \pi^+$
for pure pseudoscalar coupling is given as (see \cite{WNP}, and references therein)
\begin{equation}\epsilon \cdot M_{fi}=\sum_{j=1}^{4} A_j \overline{u}_n  
\left(\epsilon_\mu M_j^\mu \right) u_p\;\;, 
\label{eq2} 
\end{equation}
which represents an expansion based on operators
\begin{mathletters}\label{mi}
\begin{eqnarray}
M_1^\mu &=& -\gamma_5 \gamma^\mu \; k\!\!\!/ \;\;,\label{m1}\\
M_2^\mu &=& 2\gamma_5 \left( p^\mu \; k\cdot p' - p'^\mu k\cdot p \; \right)  \;\;,\label{m2}\\
M_3^\mu &=&  \gamma_5 \left( \gamma^\mu \; k\cdot p  - p^\mu  \; k\!\!\!/  \right)  \;\;,\label{m3}\\
M_4^\mu &=&  \gamma_5 \left( \gamma^\mu \; k\cdot p' - p'^\mu \; k\!\!\!/  \right)  \;\;,\label{m4}
\end{eqnarray}
\end{mathletters}
with coefficient functions
\begin{mathletters}\label{ai}
\begin{eqnarray}
A_1 &=&  \frac{ge}{s-m^2}\left( 1+\kappa_p\right) 
       + \frac{ge}{u-m^2}\kappa_n\;\;,\label{a1}\\
A_2 &=&  \frac{2ge}{(s-m^2)(t-\mu^2)}\;\;,\label{a2}\\
A_3 &=&  \frac{ge}{s-m^2}\frac{\kappa_p}{m}\;\;,\label{a3}\\
A_4 &=&  \frac{ge}{u-m^2}\frac{\kappa_n}{m}\;\;,\label{a4}
\end{eqnarray}
\end{mathletters}
where $m$ and $\mu$ are the masses of the nucleon and the pion, respectively, $g$ is the 
pseudoscalar $\pi NN$ coupling constant and $e$ the elementary
charge. The anomalous magnetic moments of 
the neutron and the proton are denoted here by $\kappa_n$ and 
$\kappa_p$. The Mandelstam 
variables $s$, $u$, and $t$ are given as (cf.\ Fig.\ \ref{fig1})
\begin{mathletters}\label{mand}
\begin{eqnarray}
s &=& (p+k)^2  = (p'+q)^2\;\;, \label{s}\\
u &=& (p'-k)^2 = (p-q)^2\;\;, \label{u}\\
t &=& (p-p')^2 = (q-k)^2\;\;, \label{t}
\end{eqnarray}
\end{mathletters}
i.e., $s+u+t=2 m^2 + \mu^2$ since all external particles are on-shell. (For
the present case, $m'=m$ in Fig.\ \ref{fig1}.)

Obviously, since each of the operators $M_i^\mu$ is gauge-invariant by itself, i.e. 
$k_\mu \cdot M_i^\mu = 0$,
the total photoproduction current is also gauge-invariant. This result obtains only if 
the vertices are bare, without any form factors. Since the terms proportional to
$M_1^\mu$, $M_3^\mu$, and $M_4^\mu$ arise from purely magnetic contributions---and 
therefore are always gauge-invariant by themselves, 
irrespective of whether one uses form factors or not---,
the problematic term, as pointed out already in Ref.\ \cite{WNP}, 
clearly is $A_2$ which arises here from the sum of the electric 
contributions of the $s$- and $t$-channels.

If one now considers the nucleons as composite objects  
and introduces form factors for the hadronic vertices,
the result for the first three diagrams of Fig.\ \ref{fig1} is
\begin{equation} \epsilon \cdot \widetilde{M}_{fi}=\sum_{j=1}^{4} \widehat{A}_j \overline{u}_n
\left(\epsilon_\mu M_j^\mu \right) u_{\rm p} + \epsilon \cdot \widetilde{M}_{\text{viol}}\;\;, 
\label{mviol}
\end{equation}
with gauge-invariant contributions
\begin{mathletters}\label{ahat}
\begin{eqnarray}
\widehat{A}_1 &=&  \frac{ge}{s-m^2}\left( 1+\kappa_{\rm p}\right) F_1 
       + \frac{ge}{u-m^2}\kappa_n F_2 \;\;,\label{ah1}\\
\widehat{A}_2 &=&  \frac{2ge}{(s-m^2)(t-\mu^2)} \widehat{F} \;\;,\label{ah2}\\
\widehat{A}_3 &=&  \frac{ge}{s-m^2}\frac{\kappa_p}{m} F_1 \;\;,\label{ah3}\\
\widehat{A}_4 &=&  \frac{ge}{u-m^2}\frac{\kappa_n}{m} F_2 \;\;,\label{ah4}
\end{eqnarray}
\end{mathletters}
and a gauge-violating term
\begin{eqnarray}
\epsilon \cdot \widetilde{M}_{\text{viol}} &=& -ge\overline{u}_{\text{n}} 
\gamma_5 \epsilon_\mu\bigg[ 
\frac{2p'^\mu}{s-m^2}(\widehat{F}-F_1)\nonumber \\
& &\quad\quad\quad\quad\quad\quad
+ \frac{2q^\mu}{t-\mu^2}(\widehat{F}-F_3)\bigg] u_{\text{p}}\;\;.\label{gviol}
\end{eqnarray}
The momentum dependence of the form factors appearing here can be read off Fig.\ \ref{fig1}, 
i.e.,
\begin{mathletters}\label{fi}
\begin{eqnarray}
F_1 &=& F_1(s) = f\!\left((p+k)^2,m'^2,\mu^2\right) \;\;, \label{f1}\\
F_2 &=& F_2(u) = f\!\left(m^2,(p'-k)^2,\mu^2\right)\;\;, \label{f2}\\
F_3 &=& F_3(t) = f\!\left(m^2,m'^2,(p-p')^2\right)  \;\;, \label{f3}
\end{eqnarray}
\end{mathletters}
(here, $m'=m$) where use is made of the fact that the form factor may always be written as a function of the
squares of the four-momenta of its three legs [cf. Eq.\ (\ref{genf})] (which does {\it not}
mean, however, that it may be taken as a function $f(s,u,t)$ of the Mandelstam
variables, as it is sometimes stated \cite{WNP}).
At this stage, $\widehat{F}$ appearing in Eqs.\ (\ref{ah2}) and (\ref{gviol}) is undefined; 
it was introduced here to be able to isolate
the gauge-violating current contribution in a form that makes comparison with Eq.\ (\ref{eq2})
easy. Clearly, the full amplitude $\epsilon \cdot \widetilde{M}_{fi}$ does not depend on it
since the sum of the $\widehat{F}$ contributions from Eq.\ (\ref{gviol}) exactly cancels 
the $\widehat{A}_2$ term.

Now, without a detailed dynamical treatment of the compositeness of
the nucleons \cite{hh97g}, {\it any} prescription for restoring gauge invariance amounts 
to introducing an additional
contact current $M_{\rm c}^\mu$ (generically depicted by the fourth diagram in Fig.\ 
\ref{fig1}),
with on-shell matrix elements cancelling exactly the gauge-violating term (\ref{gviol}), 
i.e.,
\begin{equation}
\overline{u}_n\left(\epsilon_\mu {M}_{\text{c}}^\mu\right) u_p 
     = -\epsilon \cdot \widetilde{M}_{\text{viol}}\;\;. 
\label{cc}
\end{equation}
Apart from purely transverse components or terms proportional to $k^{\mu}$, 
for the present example this contact current is
essentially given by the term in the square brackets of Eq.\ (\ref{gviol}) 
\cite{ohta89,WNP,hh97g}.
Adding this contact contribution to Eq.\ (\ref{mviol}),  one then obtains 
a gauge-invariant amplitude in analogy to Eq.\ (\ref{eq2}),
\begin{equation} \epsilon \cdot \widehat{M}_{fi}=\sum_{j=1}^{4} \widehat{A}_j \overline{u}_n
\left(\epsilon_\mu M_j^\mu \right) u_p\;\;, 
\label{mgauge}
\end{equation}
which {\it does} depend on $\widehat{F}$ now via $\widehat{A}_2$ of Eq.\ (\ref{ah2}). 

Using analytic continuation and minimal substitution, 
Ohta \cite{ohta89} finds that the required $\widehat{F}$ factor is constant,
\begin{equation}  
  \mbox{Ohta: \quad} \widehat{F} =f\!\left(m^2,m'^2,\mu^2\right)= 1\;\;,   
\label{ohtaf}
\end{equation}
determined by the normalization condition for the form factor 
in the unphysical region where all 
three legs are on-shell [see Eq.\ (\ref{genf})]. $\widehat{A}_2$ thus reduces to $A_2$ of Eq.\ (\ref{a2}), 
effectively freezing all degrees of freedom arising from the compositeness of the 
$\pi NN$ vertex and treating it like a bare one for electric current contributions. 

This determination of $\widehat{F}$ is sufficient to ensure that the
additional contact-current contribution is sin\-gu\-la\-ri\-ty-free
at $s=m^2$ and $t=\mu^2$, for 
in this limit,  both $F_1$ and $F_3$ become unity [cf. Eq.\ 
(\ref{genf})],
\begin{equation} 
F_1(s=m^2) = F_3(t=\mu^2) =1\;\;.
\label{flimit}
\end{equation}
In this limit, therefore, Eq.\ (\ref{gviol}) reduces to non-singular $\frac{0}{0}$ expressions.
Note that in the present kinematics (where all 
external particles are on-shell) one has
\begin{mathletters}
\begin{eqnarray}
s-m^2   &=& 2p \cdot k\;\;,\\
u-m'^2  &=& -2p' \cdot k\;\;,\\
t-\mu^2 &=& -2q \cdot k\;\;,
\end{eqnarray}
\end{mathletters}
and hence the limits in Eq.\ (\ref{flimit}) correspond to the vanishing of
the photon momentum. Therefore,  {\it any} (reasonably-behaved) 
subtraction function $\widehat{F}$ that becomes unity for $k=0$ 
is sufficient to restore gauge-invariance without any unwanted singularities.

In Ref.\ \cite{hh97g}, use is made of this freedom
 by allowing $\widehat{F}$ to be a function of the hadron momenta.
The only functions available that have anything to do with the physics of the present
problem are of course the form factors themselves. Haberzettl 
restores gauge invariance by constructing a contact current equivalent to choosing the 
subtraction function as
 \begin{equation}   
  \text{Choice A: \quad} \widehat{F} = F_3(t)= f\!\left(m^2,m'^2,(p-p')^2\right)\;\; 
\label{hha}
\end{equation}
which is the only function from those given in
Eqs.\ (\ref{f1})--(\ref{f3}) that does not depend explicitly on $k$ to begin with.
This, however, is an artifact of having taken both nucleon momenta as independent
variables.
Had we taken, for example, the pion momentum $q$ as an independent variable
instead of the final nucleon momentum $p'$, we would have
\begin{mathletters}\label{fiex}
\begin{eqnarray}
F_1 &=& F_1(s) = f\!\left((p+k)^2,m'^2,\mu^2\right) \;\;, \label{f1ex}\\
F_2 &=& F_2(u) = f\!\left(m^2,(p-q)^2,\mu^2\right)\;\;, \label{f2ex}\\
F_3 &=& F_3(t) = f\!\left(m^2,m'^2,(q-k)^2\right)  \;\;, \label{f3ex}
\end{eqnarray}
\end{mathletters}
which, by the same reasoning, would point to choosing $F_2$ as the subtraction function.
And if we choose $(q,p',k)$ as the independent set, we would find $F_1$.
In other words, following Ref.\ \cite{hh97g}, depending on the choice of variables, 
we can take any one of the three form factors as a subtraction function. 
In general, the subtraction vertex is the one whose single off-shell leg is described
in terms of the on-shell four-momenta of the other two legs.

One may argue whether this dependence on the variable set should be allowed. From the 
point of view of minimal substitution, however, perhaps one shouldn't find this surprising
since technically speaking, one can only perform a minimal substitution in the variables
which actually occur and hence the resulting current in general will reflect the
underlying variable set. 
Ohta circumvented this problem by considering the vertex  as a general 
function $f(p^2,p'^2,q^2)$ unconstrained by momentum conservation before performing 
the minimal substitutions. The resulting
subtraction function (\ref{ohtaf}) then corresponds to the
unphysical limit of taking all three variables to their mass-shell values.
This prescription, thus, amounts to performing the infrared limit 
$k \rightarrow 0$ explicitly in the construction of the contact current,
whereas in Ref.\ \cite{hh97g} the proper value for this limit is provided by the dynamics of
the reaction by choosing the subtraction vertex as one with
proper physical variables for its legs.
(In Ref.\ \cite{banerjee96}, some formal problems
associated with Ohta's unphysical limit have been pointed out.)

In any case, within the gauge-invariance prescription of Ref. \cite{hh97g}, 
it is possible to remove the dependence on the variable set by introducing a more
``democratic'' choice for $\widehat{F}$ using a linear combination of the three limiting 
cases, namely
\begin{eqnarray}
\text{Choice B: \quad}
\widehat{F} &=& a_1F_1(s)+a_2F_2(u)+a_3F_3(t)\nonumber\\
            &=& \widehat{F}(s,u,t)\;\;,\label{hhb}
\end{eqnarray}
where $\widehat{F}(s,u,t)$ is a short-hand notation for the preceding
expressions. To ensure the correct limit for $k=0$,
the coefficients need to add up to unity, $a_1+a_2+a_3=1$. 
The most democratic choice is $a_1=a_2=a_3= 1/3$, of course.
The previous choice A, in Eq.\ (\ref{hha}), is subsumed here with $a_1=a_2=0$, $a_3=1$.
In the subsequent applications, we will use this general form for $\widehat{F}$
and allow the coefficients $a_i$ to be free parameters.

While the equations given above for pion photoproduction apply only  at the tree level 
(in the spirit of Ref.\ \cite{WNP}), recent models have gone much further \cite{sato96,surya96,juelich96} and 
have included the pion final-state interaction by iterating the full scattering equation.  
Such a treatment would go beyond the scope of the present paper.  However, for kaon photoproducton,
most recent computations \cite{terry,david,williams} use tree-level diagrams
only and adjust the coupling constants to reproduce the data.  None of these calculations
have included a hadronic form factor until now, even though preliminary 
results \cite{bennhold96} indicate
that the presence of such a form factor greatly influences
the range of the extracted coupling constants.  We therefore 
test here this particular implementation of gauge invariance
by considering  the two kaon photoproduction reactions
$\gamma p \rightarrow \Lambda K^+$ and $\gamma p \rightarrow \Sigma^0 K^+$.

For both reactions, one can simply take over Eq.\ (\ref{ahat}) and replace the pion by $K^+$ and the
neutron by the respective hyperon. 
For $\gamma p \rightarrow \Lambda K^+$ one has
\begin{mathletters}\label{alam}
\begin{eqnarray}
\widehat{A}_{\Lambda 1} &=&  \frac{g_\Lambda e}{s-m^2}\left( 1+\kappa_p\right) F_{\Lambda 1}(s) 
\nonumber\\
       & &\quad + \frac{g_\Lambda e}{u-m^2_\Lambda}\kappa_\Lambda F_{\Lambda 2}(u) \;\;,\label{alam1}\\
\widehat{A}_{\Lambda 2} &=&  \frac{2g_\Lambda e}{(s-m^2)(t-\mu^2)} \widehat{F}_{\Lambda}(s,u,t) \;\;,\label{alam2}\\
\widehat{A}_{\Lambda 3} &=&  \frac{g_\Lambda e}{s-m^2}\frac{\kappa_p}{m} F_{\Lambda 1}(s) \;\;,\label{alam3}\\
\widehat{A}_{\Lambda 4} &=&  \frac{g_\Lambda e}{u-m^2_\Lambda}\frac{\kappa_\Lambda}{m_\Lambda} F_{\Lambda 2}(u) \;\;,\label{alam4}
\end{eqnarray}
\end{mathletters}
where $F_\Lambda$ is the $\Lambda K p$ form factor, with coupling constant 
$g_{K\Lambda N}$, and $m_\Lambda$ is the $\Lambda$ mass; $\kappa_\Lambda$ is the corresponding 
anomalous magnetic moment.
For the second reaction, $\gamma p \rightarrow \Sigma^0 K^+$, one replaces  
$\Lambda$ by $\Sigma$.

Clearly, a phenomenological description of the $(\gamma, K)$ processes has to include 
resonance terms. However, the quality of the data has not yet
permitted a clear identification of the relevant resonances in the reaction
mechanism and, consequently, models with different resonances can all achieve
a satisfactory description of the data \cite{terry,david,williams}.
These resonance terms are all gauge-invariant independently and, therefore,
do not depend on different prescriptions of restoring gauge invariance.  For our
empirical studies below we choose the same set of resonances as
in Refs.\ \cite{terry,bennhold96}, namely, the $K^*$ in the $t$-channel, and
the $S_{11}(1650)$ and the $P_{11}(1710)$ states in the $s$-channel.
For $\Sigma$ production, we also allow the 
$S_{31}(1900)$ and the $P_{31}(1910)$ state to contribute.
 We do not make any claims that this selection is unique
or correct at the present time, but rather that it leads to a reasonable description
of the $(\gamma, K)$ processes and allows us to draw qualitative conclusions
about the magnitude of the Born coupling constants.  In the case of 
$p(\gamma,K^+)\Lambda$, separate coupled-channels
 analyses\cite{manley,feuster97}
 found the $S_{11}(1650)$ and the $P_{11}(1710)$ states to play important roles.
For simplicity, all resonances are multiplied here with the same hadronic form factor.

For the numerical evaluation of Eqs.\ (\ref{alam}), we choose covariant vertex 
parameterizations without any singularities on the real axis. 
For a baryon with mass $m$ and four-momentum $p$ decaying (virtually)
into a baryon with mass $m'$ and four-momentum $p'$ and a meson with mass $\mu$ 
and momentum $p-p'$, 
the general vertex may be written as
\begin{equation}
    F = f\!\left( p^2,p'^2,(p-p')^2 \right) \;\;,
\label{genf}
\end{equation}
with the normalization $f\!\left( m^2, m'^2, \mu^2  \right)=1$. 
When applied to 
$\gamma p \rightarrow \Lambda K^+$ and $\gamma p \rightarrow \Sigma^0 K^+$,
the masses $m$  and $\mu$ appearing in Eq.\ (\ref{fi})
are always the nucleon and kaon masses, respectively,
whereas $m'=m_\Lambda$ for the first and $m'=m_\Sigma$ for the second reaction. 
The vertex parameterization we employ here is of the form
\begin{equation}
f \!\left( p'^2,p^2,(p-p')^2 \right) = \frac{\Lambda^4}{\Lambda^4+\eta^4}\;\;,\label{fgen}
\end{equation}
where $\Lambda$ is some cutoff parameter, and
\begin{eqnarray}
\eta^4 &=&\left(  p^2    - m^2  \right)^2
               +\left( p'^2    - m'^2 \right)^2  \nonumber\\
       & & \quad\quad +\left((p-p')^2 -\mu^2 \right)^2\;.
\label{ch1}
\end{eqnarray}
In the nonrelativistic 
limit, this form reduces to the usual monopole form depending on the 
squared three-momentum of the exchanged particle.
For the three cases of Eq.\ (\ref{fi}), since two of the three vertex legs are 
always on-shell in the present applications, this translates into
\begin{mathletters}\label{ffchoice1}
\begin{eqnarray}
F_1 &=& \frac{\Lambda^4}{\Lambda^4+\left( s-m^2 \right)^2} \;\;,\\
F_2 &=& \frac{\Lambda^4}{\Lambda^4+\left( u-m'^2 \right)^2} \;\;,\\
F_3 &=& \frac{\Lambda^4}{\Lambda^4+\left( t-\mu^2 \right)^2} \;\;,
\end{eqnarray}
\end{mathletters}
which is, therefore, effectively the same as the form factors used in
Ref. \cite{ffparam}. 

In the discussion of our numerical results, we focus our attention on the magnitude of the
leading Born coupling constants $g_{K\Lambda N}$ and $g_{K\Sigma N}$.
In contrast to the well-known $\pi NN$ coupling constant,
there are  serious discrepancies between
values for the {\it KYN}
coupling constants extracted from electromagnetic 
reactions \cite{williams,bennhold96}
and those from hadronic processes \cite{anto,timmer} 
which tend to be closer to accepted SU(3) values.
If the leading coupling constants $g_{K \Lambda N} / \sqrt{4 \pi}$ and
$g_{K \Sigma N} / \sqrt{4 \pi}$ are not allowed to vary freely and are fixed
(close to what is obtained from hadronic reactions \cite{timmer}) 
at reasonable SU(3) values of $-3.8$ and 1.2, respectively, 
the $\chi^2$ obtained in our model {\it without} hadronic form factors
for the $(\gamma, K)$ reactions comes out to be 55.8.  
If, on the other hand, the two couplings are allowed
to vary freely, one obtains $g_{K \Lambda N} / \sqrt{4 \pi} = -1.89$ and
$g_{K \Sigma N} / \sqrt{4 \pi} = -0.37$.
This clearly indicates that either
there is a very large amount of SU(3) symmetry breaking or that important
physics has been left out
in the extraction of coupling constants from the $(\gamma,K)$ processes.
In this study, we advocate the second position and demonstrate that 
the inclusion of structure at the hadronic vertex permits an adequate
description of kaon photoproduction with couplings close to the
SU(3) values, provided one uses the gauge procedure
of Ref.\ \cite{hh97g}.

\begin{figure}[t!]
\centerline{\psfig{file=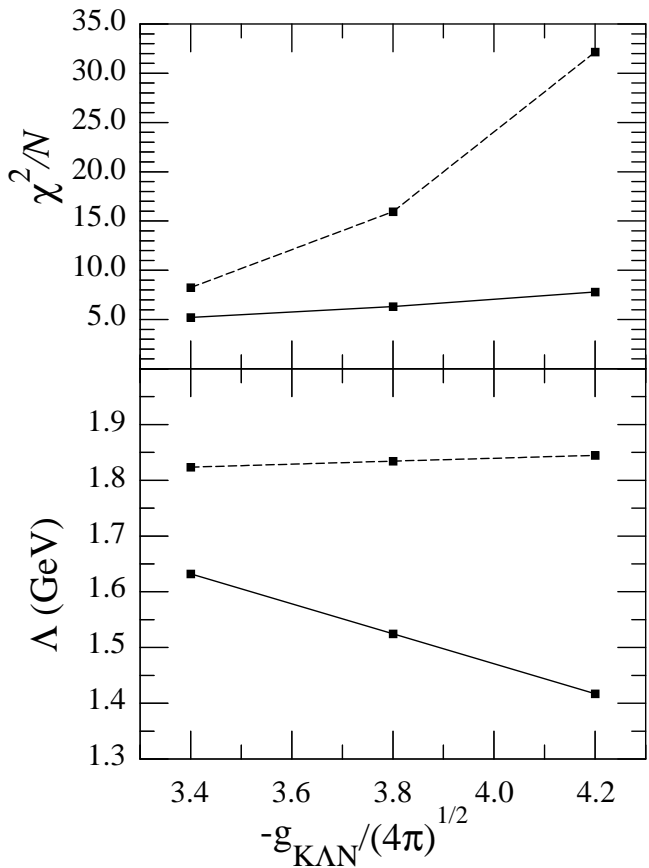,width=0.99\columnwidth,clip=,silent=}}
\vspace{3mm}
\caption[title]{\label{fig2} Values of $\chi^2/N$ (where $N$ is the number of data points)
and cutoff parameter $\Lambda$ for
coupling constant values of $-g_{K\Lambda N} / \sqrt{4 \pi}=3.4$, 3.8, and 4.2. The solid
lines connect results obtained with Haberzettl's gauge formalism \cite{hh97g} and the dotted 
lines pertain to Ohta's \cite{ohta89} prescription. }
\vspace{-4mm}
\end{figure}

The main numerical results of our investigation are summarized in 
Fig.\ \ref{fig2}. 
%
%
%
%
The upper panel shows $\chi^2$ per data point as a function of
$g_{K \Lambda N} / \sqrt{4 \pi}$ for the two 
different gauge prescriptions by
Ohta and Haberzettl.  At a value of $g_{K \Lambda N} / \sqrt{4 \pi}=-3.4$, the $\chi^2$
obtained with Ohta's method is almost a factor 2 larger compared to using
the method by Haberzettl. With increasing coupling  constant the Ohta result
rises sharply, leading to an unacceptably large $\chi^2$ of 32.2 for 
$g_{K \Lambda N} / \sqrt{4 \pi} = -4.2$.   On the other hand, using the 
procedure of Ref.\ \cite{hh97g} keeps the $\chi^2$ more or less constant.
This dramatic difference between the two gauge prescriptions can easily be
understood from Eq.\ (\ref{ohtaf}) and the discussion following that equation.  
Ohta's method provides no possibility
to suppress electric contributions since the form factor for this 
term is unity [cf. Eqs.\ (\ref{ah2}) and (\ref{ohtaf})].
In contrast, the method by Haberzettl allows for a  hadronic form factor
in this term as well.

The lower panel of Fig.\ \ref{fig2} sheds additional light on the 
suppression mechanism. 
In the fits we performed the cutoff $\Lambda$
of the form factor, cf.\ Eq.\ (\ref{fgen}),  was allowed
to vary freely. In the case of
Haberzettl's method, the cutoff decreases with
increasing $K \Lambda N$ coupling constant, leaving the magnitude of the
{\it effective} coupling, i.e., coupling
constant times form factor,  roughly constant. 
Again, since Ohta's method
does not involve form factors for electric contributions no such
compensation is possible there, and as a consequence
the cutoff remains insensitive to the coupling constant.

In obtaining Fig.\ \ref{fig2} we have kept $g_{K \Sigma N}$ fixed at the
value $g_{K \Sigma N} / \sqrt{4 \pi} = 1.2$. We have checked that varying the
$K \Sigma N$ coupling between 1.0 and 1.4 leads only to very small changes.
Furthermore, we allowed the coefficients $a_i$ of Eq.\ (\ref{hhb}) to be free
fit parameters. As it turns out, the fit only allows nonzero $s$- and $t$-channel
contributions (i.e., $a_2$ is essentially zero), with a somewhat larger $a_3$ value 
(corresponding to an enhancement of the $t$-channel), which of course is entirely
consistent with the fact that Eq.\ (\ref{alam2}) contains only $s$- and $t$-channels.

We do not show the fitted resonance couplings here since we do not regard
them as very realistic at this point.
We emphasize again the qualitative nature of our findings, and
clearly a more sophisticated calculation is required in order to obtain
a quantitative description of the $(\gamma, K)$ processes.

In summary, we have applied here the general gauge-invariance restoration method
proposed by Haberzettl to the specific example of
pseudoscalar photoproduction at the tree level.  Using
a phenomenological Born plus resonance model we have compared
the procedures by Ohta \cite{ohta89} and Haberzettl \cite{hh97g} for kaon photoproduction. 
We found the latter to be superior since it 
can provide a resonable description of the
data using values for the leading couplings constants close to the SU(3) values.
Such couplings cannot be accommodated in Ohta's method due to the 
absence of a hadronic form factor in the electric current contribution.
The main purpose for measuring meson photoproduction
in the 1--2 GeV region is the study of resonances.
In order to unambiguously separate resonance
from background contributions, it is imperative
that  background terms be able to account
for hadronic structure while properly maintaining
gauge invariance. As the present findings indicate, Ohta's prescription seems to be too
restrictive in this respect, whereas the method put forward in Ref.\ \cite{hh97g} seems
well capable of providing this facility.

This work was supported in part by Grant No.\ DE-FG02-95ER40907 of the U.S. Department of 
Energy.


\vspace{-4mm}
\begin{references}
\vspace{-14mm}
\bibitem{gaugeauthors}  
F. Gross and D. O. Riska, 
  Phys.\ Rev.\ C{\bf 36}, 1928 (1987);
H. W. L. Naus, J. H. Koch, and J. L. Friar, 
  Phys.\ Rev.\ C {\bf 41}, 2852 (1990);
J. W. Bos, S. Scherer, J. H. Koch,
  Nucl.\ Phys.\ {\bf A547}, 488 (1992).

\bibitem{ohta89}
K. Ohta, 
  Phys. Rev.\ C {\bf 40}, 1335 (1989).

\bibitem{WNP}
R. L. Workman, H. W. L. Naus, S. J. Pollock, 
  Phys.\ Rev.\ C {\bf 45}, 2511 (1992). 

\bibitem{hh97g}
H. Haberzettl, 
  Phys.\ Rev.\ C {\bf 56}, 2041 (1997).

\bibitem{fmult}
J. C. Bergstrom, 
  Phys.\ Rev.\ C {\bf 44}, 1768 (1991);
J. Cohen, 
  Int.\ J. Mod.\ Phys.\ A {\bf 4}, 1 (1989);
S. Nozawa, B. Blankleider, and T.-S. H. Lee, 
  Nucl.\ Phys.\ {\bf A513}, 459 (1990).

\bibitem{banerjee96}  
S. Wang and M. K. Banerjee, 
  Phys.\ Rev.\ C {\bf 54}, 2883 (1996).

\bibitem{sato96} 
H.\ Sato and T.-H. S.\ Lee, 
  Phys.\ Rev.\ C{\bf 54}, 2660 (1996).


\bibitem{surya96}
Y. Surya and F. Gross, 
  Phys.\ Rev.\ C {\bf 53}, 2422 (1996).

\bibitem{juelich96}
K. Nakayama {\it et al.}, 
  {\it Proceedings of the 4th CEBAF/INT Workshop on $N^*$ Physics}, 
  INT, Seattle, Sept. 9-13, 1996 (World Scientific, Singapore, 1997, T.-S.H. Lee and
  W. Roberts, editors), p.\ 156.


\bibitem{terry} 
T. Mart, C. Bennhold, and C. E. Hyde-Wright, 
  Phys.\ Rev.\ C {\bf 51}, R1074 (1995).

\bibitem{david} 
J. C. David, C. Fayard, G. H. Lamot, and B. Saghai,
  Phys.\ Rev.\ C {\bf 53}, 2613 (1996).

\bibitem{williams} 
R. A. Williams, C.-R. Ji, and S. R. Cotanch,
  Phys.\ Rev.\ C {\bf 46}, 1617 (1992).

\bibitem{bennhold96}
C. Bennhold, T. Mart, and D. Kusno, 
  {\it Proceedings of the 4th CEBAF/INT Workshop on $N^*$ Physics}, 
  INT, Seattle, Sept. 9-13, 1996 (World Scientific, Singapore, 1997, T.-S. H. Lee and
  W. Roberts, editors), p.\ 166; 
T. Mart and C. Bennhold, 
  Few-Body Systems Suppl.\ {\bf 9}, 213 (1995).

\bibitem{manley} 
D. M. Manley and E. M. Saleski, 
  Phys.\ Rev.\ D {\bf 45}, 4002 (1992).

\bibitem{feuster97}
T. Feuster and U. Mosel, 
  Los Alamos eprint nucl-th/9708051 (1997).

\bibitem{ffparam}
B. C. Pearce and B. K. Jennings, 
  Nucl.\ Phys.\ {\bf A528}, 655 (1991).

\bibitem{anto} 
J. Antolin, 
  Z. Phys.\ {\bf C31}, 417 (1986).

\bibitem{timmer} 
R. G. E. Timmermans {\it et al.}, 
  Nucl. Phys.\ {\bf A585}, 143c (1995).

\end{references}
\end{document}